\documentclass[aps,pra,superscriptaddress,showkeys,twocolumn]{revtex4-1}
\usepackage{amsfonts,amsmath}
\usepackage{epsfig}
\usepackage{bm}
\usepackage{amsmath}
\usepackage{subfigure}
\usepackage{graphicx}
\usepackage{braket}

\newcommand{\norm}[1]{\left\lVert#1\right\rVert}

\newcommand{\beq}{\begin{equation}}
\newcommand{\eeq}{\end{equation}}
\newcommand{\bea}{\begin{eqnarray}}
\newcommand{\eea}{\end{eqnarray}}

\newcommand{\rmd}{{\text d}}

\usepackage{mathtools}

\usepackage{graphicx,amsmath}
\usepackage{color}

\begin{document}

\title{Realizable Quantum Adiabatic Search}
\author{Itay Hen}
\email{itayhen@isi.edu}
\affiliation{Information Sciences Institute, University of Southern California, Marina del Rey, CA 90292, USA}
\affiliation{Department of Physics and Astronomy and Center for Quantum Information Science \& Technology,University of Southern California, Los Angeles, California 90089, USA}

\begin{abstract}
Grover's unstructured search algorithm is one of the best examples to date for the superiority of quantum algorithms over classical ones. Its applicability, however, has been questioned by many due to its oracular nature. We propose a mechanism to carry out a quantum adiabatic variant of Grover's search algorithm using a single bosonic particle placed in an optical lattice.  By studying the scaling of the gap and relevant matrix element in various spatial dimensions, we show that a quantum speedup can already be gained in three dimensions. We argue that the suggested scheme is realizable with present-day experimental capabilities. 
\end{abstract}

\keywords{adiabatic quantum computing, spatial search, Grover's algorithm, Bose-Hubbard model} 
\maketitle

\section{Introduction}

Along with Shor's polynomial-time algorithm for integer factorization~\cite{shor:94}, 
Grover's algorithm for the speedy search of an unstructured database~\cite{grover:97} is considered to be the tour de force of quantum computing,  exhibiting the best example to date of the superiority of quantum computers over classical ones.
The exciting possibility that quantum computers will actually be able to quickly retrieve items by quantum mechanically sifting through databases serves as a powerful driving force for both theoretical and experimental research in the field of quantum computing. 

Unlike Shor's integer factorization, Grover's algorithm is a `black-box' routine that assumes the existence of an omniscient quantum oracle capable of responding to queries instantaneously, regardless of the size of the database. This at least seemingly problematic requirement has raised doubts as to the physical applicability of the algorithm. First to recognize this matter was Benioff~\cite{benioff:02} who studied the question of whether Grover's algorithm can speed up the search of a physical region, 
noting that when searching a two-dimensional grid with $N$ sites, the algorithm must use on the order of $O(\sqrt{N})$ steps to return to its starting point during each of the $\sqrt{N}$ Grover iterations. Refuting Benioff's assertion, Aaronson and Ambainis~\cite{spatialSearchAaronson} later pointed to other fundamental physical limits placed on information storage, which stem from the holographic principle of black hole thermodynamics. 

Analogs of Grover's unstructured search algorithm that yield quadratic speedups have also been devised in the framework of adiabatic quantum computing (AQC)~\cite{roland:02,hen:14,hen:14b}. Here too, the physical realizability of the adiabatic oracle has been called into question due to the highly non-local nature of the Hamiltonian, which also consists of exponentially many terms.

In what follows, we study the computational power of physically realizable quantum adiabatic search processes.
Specifically, we consider a bosonic particle that is allowed to adiabatically hop between neighboring sites of an optical lattice.  We will analytically show that in (the unrealistic case of) four and higher dimensions, an adiabatic spatial search by a quantum particle can be quadratically faster than the corresponding classical search. Additionally, by numerically studying the scaling of the gap and relevant matrix element with problem size, we will show that by applying a carefully chosen adiabatic schedule, quantum speedup may already be attained in three dimensions (but not in two). As we also argue, such a quantum speedup can, under certain conditions, be demonstrated experimentally in a lab using currently available technology.

\section{Preliminaries} AQC~\cite{Finnila1994343,Brooke30041999,kadowaki:98,farhi:01,santoro:02} is a paradigm of computing that utilizes gradually decreasing quantum fluctuations to find the global optima of discrete optimization problems~\cite{young:08,young:10,hen:11,hen:12,farhi:12}. 
In AQC, the solution to an 
optimization problem is encoded in the ground state of a problem Hamiltonian $H_p$ that defines a cost function whose minimum is sought.
To reach a minimizing configuration of $H_p$, the system is initially prepared in the ground state of another `driver' Hamiltonian
$H_d$, chosen so that it does not commute with $H_p$ and has a ground state that is easy to prepare. 
The total Hamiltonian of the system then slowly interpolates between $H_d$ to
$H_p$ via, e.g., 
\beq\label{eq:hs}
{H}(s)=(1-s) H_d + s H_p \,,
\eeq
where $s(\tau)$ is a parameter varying smoothly with time $\tau$
from $s(0)=0$ initially to $s(\mathcal{T})=1$ at the end of the evolution, at which point the quantum fluctuations generated by $H_d$ vanish. 
If this process is done slowly enough, the
adiabatic theorem of quantum mechanics~\cite{kato:51,jansen:07,lidarGap,Elgart}
ensures that the system will stay close to the ground state of the
instantaneous Hamiltonian throughout the evolution, so that one finally
obtains a state close to the ground state of $H_p$.   
The running time $\mathcal{T}$ of the algorithm determines the
efficiency, or complexity, of the algorithm.
A generic  condition
 for the adiabatic approximation to hold
can be given in terms of the instantaneous eigenstates $\{ | m
\rangle \}$ and eigenvalues $\{E_m \}$ of the Hamiltonian $H(s)$,
as~\cite{wannier:65,farhi:02}
\begin{equation} \label{eq:T}
\mathcal{T} \geq \frac1{\epsilon} \, {\max_{s} V_{01}(s)  \over
\min_s g^2(s)} \,,
\end{equation} 
where $g(s)$ is the first
excitation gap $E_1(s)-E_0(s)$ and $V_{01}(s) = \left| \langle 0 | \rmd H / \rmd s| 1\rangle \right|$ (in our units $\hbar=1$).
Here, $\epsilon$ is a small number inversely proportional to the running time of the algorithm. While for certain Hamiltonians more stringent conditions may be required (see, e.g., Ref.~\cite{jansen:07} and references therein), for the systems we shall consider here, the above inequality will suffice; the time-dependence of the parameter $s$ on time $\tau$ we will consider is smoothly varying, a condition that has been shown to suffice in order for the above bound to hold~\cite{Elgart}.

In adiabatic unstructured search problems, the cost function encoded in $H_p$ is constant across the entire search space (which corresponds to the diagonal elements of the Hamiltonian) except for a limited set of `marked' configurations whose cost is lower than the rest.
These constitute the solution space. Roland and Cerf~\cite{roland:02} demonstrated that if both problem and driver Hamiltonians are encoded by one-dimensional projections, $H_p$ projecting onto the marked state and $H_d$ onto the equal superposition of all computational basis states,
a quantum adiabatic algorithm applied to Grover's problem results in a quadratic speedup.
To achieve a speedup, one must carefully choose a variable annealing schedule, a principle commonly referred to as `local adiabatic evolution' (LAE), wherein a local Landau-Zener condition must be satisfied locally~\cite{jansen:07,roland:02}, namely,
\beq \label{eq:lae}
\left|\frac{\rmd \tau}{\rmd s}\right| \geq \frac1{\epsilon}  \frac{V_{01}(s)}{g^2(s)}\,.
\eeq 
As noted above, the use of one-dimensional projections to encode the Hamiltonian of the system renders the problem physically unrealizable due to the Hamiltonian being highly non-local (explicitly, $n$-local where $n$ is the number of spins), as well as consisting of exponentially many $k$-body terms that are essentially impossible to realize in practice\footnote{In terms of distinct  $k$-body terms, the $n$-qubit one-dimensional projection onto the $\ket{00\ldots0}$ state decomposes to a sum of $2^n$ terms:
$\ket{00\ldots0}\bra{00\ldots 0} \propto 1+\sum_i \sigma^z_i +\sum_{i \neq j} \sigma^z_i \sigma^z_j + \sum_{i \neq j \neq k} \sigma^z_i \sigma^z_j \sigma^z_k + \ldots + \prod_i \sigma^z_i$ with a similar decomposition for other one-dimensional projections.}. 
Here, we consider a different setting for an adiabatic process describing a quantum unstructured search---one that is physically more meaningful and in which the above complications do not arise. 

\section{Searching an optical lattice with a single boson}
 Let us consider a gas of bosons placed in a $d$-dimensional optical trap consisting of a periodic optical potential with $N=L^d$ sites.
The physics of this model is given by the so-called Bose-Hubbard Hamiltonian~\cite{jaksch}
\beq
 \label{eq:Ham}
{H} = - t \sum_{\langle ij \rangle} \left( {a}_i^{\dagger} {a}_j 
+ {a}_j^{\dagger} {a}_i \right)  +\frac{U}{2} \sum_i {n}_i({n}_i-1) - \sum_i \mu_i {n}_i \,. \nonumber\\
\eeq
The Bose-Hubbard model exhibits a phase transition at zero temperature from a superfluid to a Mott-insulating phase, forming one of the paradigm examples
of a quantum phase transition~\cite{sachdev:99}. Here, $\langle ij \rangle$ denotes nearest neighbors, ${a}_i$ (${a}_i^{\dagger}$) 
destroys (creates) a boson on site $i$, ${n}_i={a}_i^{\dagger} {a}_i$ 
is the local density operator,  and $\mu_i$ denotes a local chemical potential. 
The hopping parameter $t>0$ sets the energy scale, and $U$ is the strength of the onsite repulsion potential. For simplicity, we shall consider here the limit of large $U$, also known as the `hardcore' limit, where the onsite repulsion is so strong that the boson creation and annihilation operators satisfy the constraints ${a}^{\dagger 2}_{i}= {a}^2_{i}=0$ and 
$\left\{ {a}_{i},{a}^{\dagger}_{i} \right\}=1$
which prohibit double or higher occupancy of lattice sites. The Bose-Hubbard Hamiltonian has been realized in a lab 
in one, two and three dimensions~\cite{bloch,Spielman,Spielman2,greiner}. 

We next set up our to-be-optimized cost function by specifying the values of the local chemical potential $\mu_i$. 
We set these to
\beq
\mu_i=\Big\{
\begin{tabular}{rl}
$\mu$ & \text{if} \, $i=i_*$\\
$0$ &  \text{otherwise}\\
  \end{tabular}
\eeq
for some fixed $\mu>0$ and an unknown site index $i_*$ that is to be found.  Since the cost function prescribed by the local chemical potential is completely unstructured, a classical particle searching through the lattice will find the marked site in $O(N)=O(L^d)$ steps on average (and as a worst-case scenario). In contrast to computational search in which the search space grows exponentially with the number of quantum bits, here the search space corresponds directly to the $N$ lattice sites.
In what follows we demonstrate that a hardcore boson, allowed to hop between neighboring lattice sites, can find the marked site faster than its classical counterpart if the gradual adiabatic turning off of the hopping term and simultaneous turning on of the chemical potential are allowed.

\begin{figure*}[htp]
{\includegraphics[width=0.32\textwidth]{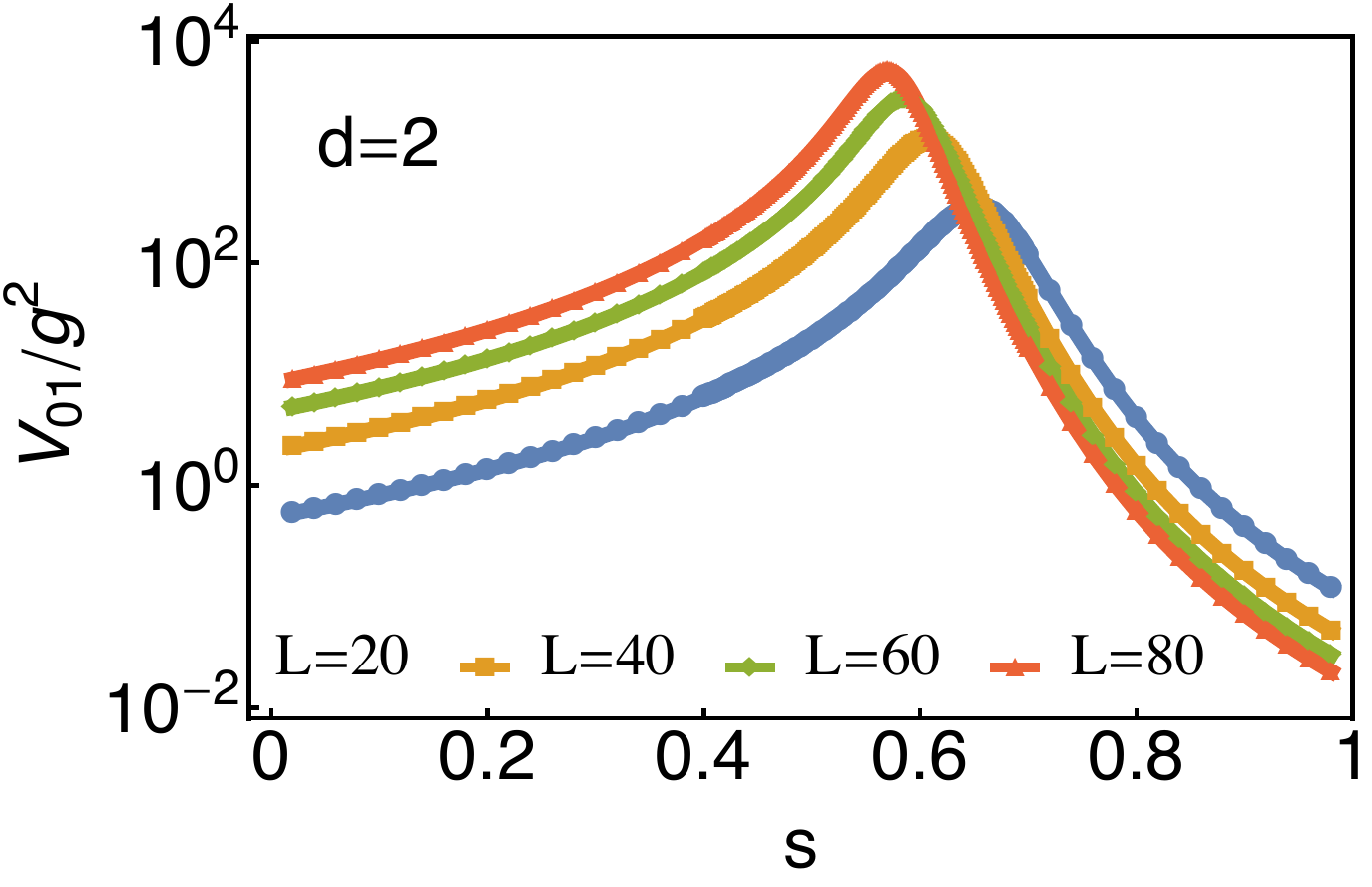}}
{\includegraphics[width=0.32\textwidth]{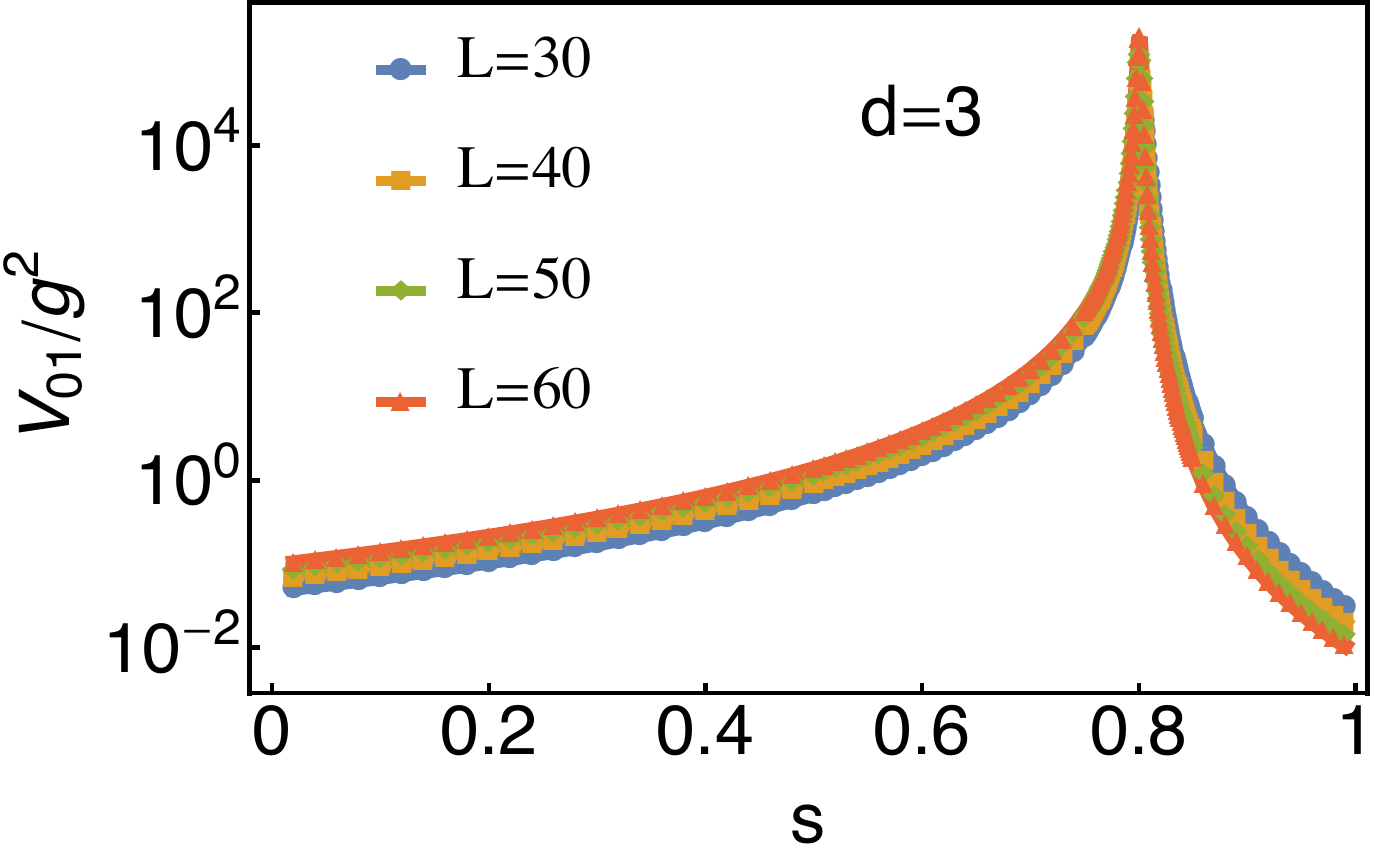}}
{\includegraphics[width=0.32\textwidth]{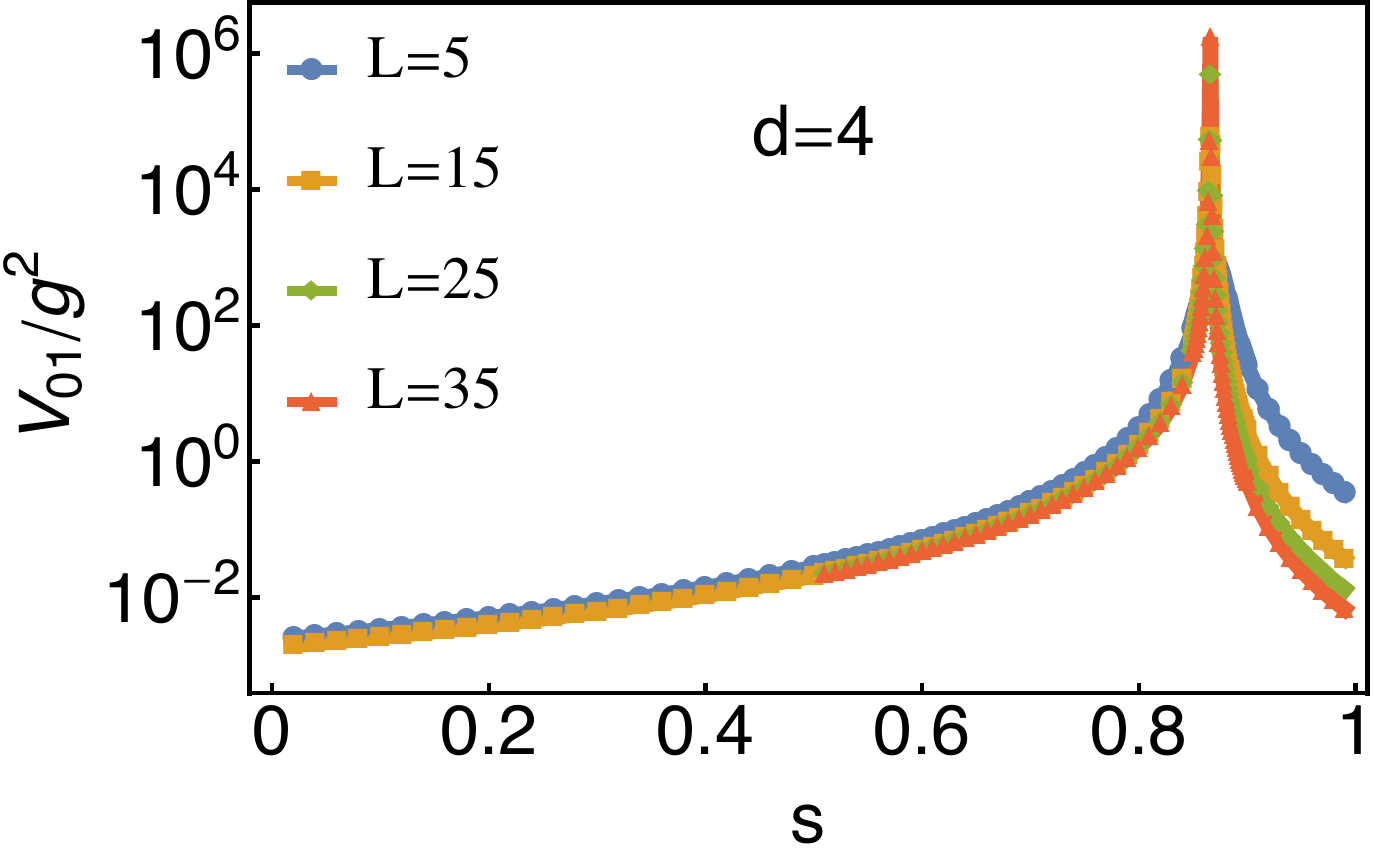}}
\caption{{\bf The integrand $|\rmd \tau/\rmd s|=V_{01}(s)/g^2(s)$ throughout the annealing for different problem sizes $L=N^{1/d}$ in dimensions $d=2,3$ and $4$ (left to right).}  All curves exhibit the `usual' peak corresponding to a closing minimum gap. The area under the curves corresponds to the total runtime $\mathcal{T}$ of the adiabatic process. In this logarithmic plot, we see that the contribution to $\mathcal{T}$ comes mainly from the vicinity of the peak.}
\label{fig:dtdg}
\end{figure*}

For notational convenience, we shall henceforth use the fact that the Bose-Hubbard model may be viewed in the hardcore limit
as an $XY$ model of a spin-$1/2$ system. This correspondence is provided by the Holstein-Primakoff transformation between bosonic operators and $SU(2)$ generators, explicitly,
\bea
{a}_i^{\dagger} &\leftrightarrow &\sigma_i^{+}/2 \quad \text{and}\quad 
{a}_i \leftrightarrow \sigma_i^{-}/2 \,,
\eea
in which case the local density operator ${n}_i={a}_i^{\dagger} {a}_i$ corresponds to $(1+\sigma_i^{z})/2$.
With the above mapping, the hardcore boson Hamiltonian becomes
an $XY$ antiferromagnet with a magnetic field applied to the marked site:
\beq\label{eq:Hb}
H(s)=-(1-s) t \sum_{\langle i j \rangle} \left(\sigma^x_i \sigma^x_j+\sigma^y_i \sigma^y_j\right) - \frac1{2} s \mu \left(1+\sigma^z_{i_*}\right)  \,,
\eeq
where $s(\tau)$ is a tunable adiabatic parameter smoothly varying from $0$ to $1$ throughout the evolution. 
The Bose-Hubbard Hamiltonian conserves the number of particles throughout the evolution~\cite{CQA1,CQA2}. In spin terminology, this corresponds to conservation of total $z$-magnetization. In the one-particle sector considered here, the evolution of a state will be restricted to the $N$-dimensional subspace spanned by the $N$ basis states with exactly one spin pointing up:
\beq
|i\rangle \equiv |\downarrow_1 \downarrow_2 \cdots \uparrow_i  \cdots \downarrow_N\rangle \quad \text{with} \quad i=1\ldots N \,.
\eeq
In this subspace, the Hamiltonian is reduced to
\beq\label{eq:Hred}
H(s)=-(1-s) t A- s \mu  |i_*\rangle \langle i_*|  \,,
\eeq
where 
\beq
A=\sum_{\langle i j \rangle} \left( |i\rangle \langle j|+|j\rangle \langle i|\right)
\eeq
is the adjacency matrix of the periodic lattice. 
Because of the lattice symmetries, the spectrum of the Hamiltonian, and as a consequence    the eventual overall performance of the adiabatic process, can be determined independently of the label $i_*$ of the marked site, and so one may derive the complexity of the quantum adiabatic algorithm regardless of the choice of $i_*$. 

At $\tau=0$, where $s=0$, the ground state of the Hamiltonian is simply the equal superposition
\hbox{$|+\rangle = \frac1{\sqrt{N}}|i\rangle$} (with energy $E_0=-2 d t$~\cite{PhysRevLett.105.180401}) corresponding to a fully delocalized particle. At the end of the evolution when $s=1$,
the ground state approaches the fully localized $|i_*\rangle$ (with energy $E_0 = -\mu$).
Let us now estimate the runtime $\mathcal{T}$ of an adiabatic transition from the initial ground state $|+\rangle$ at $\tau=0$ to the final $|i_*\rangle$ at $\tau=\mathcal{T}$. We will do so by placing bounds on the gap $g(s)$ and the matrix element $V_{01}(s)$ as per the conditions given in eqs.~(\ref{eq:T}) and~(\ref{eq:lae}). 

\section{Analytically derived bounds} The observant reader will notice that for any fixed value of $s$, the Hamiltonian, eq.~(\ref{eq:Hred}), depicts in fact a spatial search by a (continuous-time) quantum random walker on a $d$-dimensional cubic graph.  This system has been examined by Childs and Goldstone~\cite{spatialSearchChilds} who analyzed its spectrum, showing that in dimensions four and higher, the minimum gap of the system $\min_s g(s)$ scales in the large $N$ limit as $O(1/\sqrt{N})$ and in a region of $s$ of order \hbox{$\mathcal{W} \sim 1/\sqrt{N}$} (where in $d=4$ there are additional logarithmic corrections). 

Additionally, the matrix element $V_{01}(s)$ can be bound by
\bea
V_{01}(s) &=&\left| \langle 0| \rmd H/\rmd s |1\rangle\right|= \left| \mu \langle 0 |i_*\rangle \langle i_*|1\rangle -t \langle 0| A |1\rangle\right| \nonumber\\ 
&\leq& \mu \norm{ i_*\rangle \langle i_*}+ t \norm{A} =\mu + 2 d t \,,
\eea
that is, $\max_s V_{01}(s)$ can be bound by a constant that does not depend on lattice size (and only linearly on dimension)\footnote{The above bound can actually be made tighter but will suffice for the current purposes.}. 
Combining the above two bounds, the overall runtime for a constant-rate adiabatic evolution for a hardcore boson in a $d\geq4$-dimensional periodic lattice can thus be approximated as:
\beq
\mathcal{T} \propto \frac{\max_{s} V_{01}(s)}{\min_{s}g^2(s) } \sim O(N)\,,
\eeq
 i.e., scaling linearly with problem size, similar to its classical counterpart. 
 
However, since the gap and matrix element of the problem at hand are calculable for all values of $s$ independently of $i_*$, one may take advantage of LAE---that is, obtain further speedups by choosing a variable-rate schedule which slows down in the vicinity of the minimal gap but speeds up when the gap is large. 
Using the fact that the gap is small in $d \geq 4$ only over a small region $\mathcal{W}\sim 1/\sqrt{N}, $ one can apply local adiabatic evolution which immediately yields the usual quadratic speedup
\beq
\mathcal{T} \propto \mathcal{W} \frac{\max_{s} V_{01}(s)}{\min_{s} g^2(s)} \sim O(\sqrt{N}) \,.
\eeq
\\
\section{Numerical analysis} 
Beyond the analytically derivable bounds discussed above for dimensions $d\geq4$, it is of particular interest to study the scaling of the runtime in lower dimensions, in which case the Bose-Hubbard Hamiltonian can be physically realized. 
In order to calculate the scaling of the LAE runtime~\cite{roland:02}, given by
\beq
\mathcal{T} =\int \rmd s \left| \frac{\rmd \tau}{\rmd s}\right| =\int \rmd s \frac{V_{01}(s)}{g^2(s)}\,,
\eeq
we study the behavior of the gap $g(s)$ and matrix element $V_{01}(s)$ as a function of problem size. 
The numerically evaluated integrand \hbox{$|\rmd \tau/\rmd s|=V_{01}(s)/g^2(s)$} is plotted in fig.~\ref{fig:dtdg} for dimensions $d=2,3$ and $4$ (the model parameters are fixed for simplicity at $t=1$ and $\mu=1$). As one might expect, the quantity $|\rmd \tau/\rmd s|$ becomes more sharply peaked with increasing system size around a critical value of $s$  where both the gap attains its minimum and the matrix element its maximum.

\begin{figure}[htp] 
   \centering
   \includegraphics[width=0.95\columnwidth]{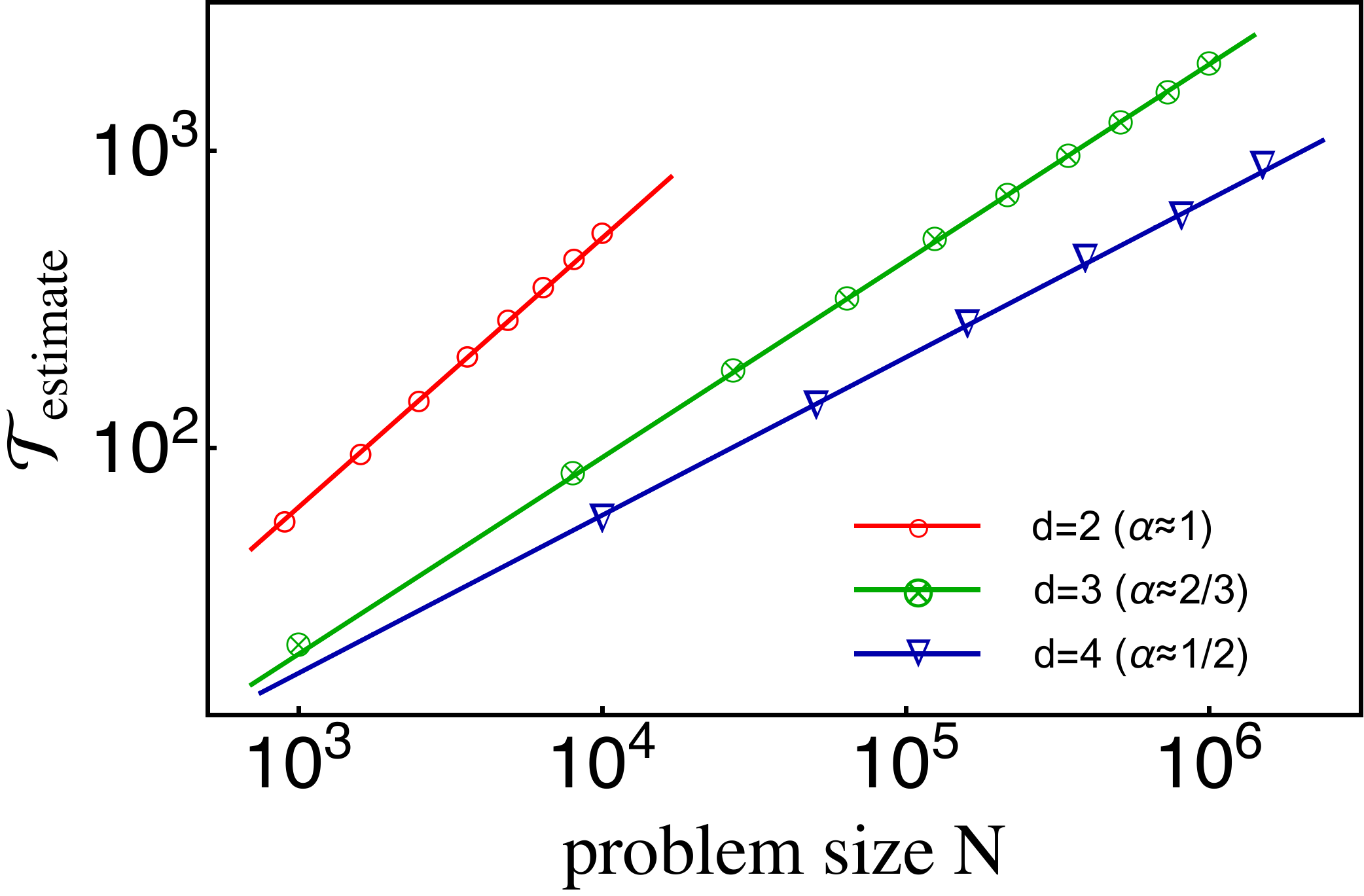} 
   \caption{{\bf Numerical scaling of runtime with problem size $N=L^{d}$ using local adiabatic evolution:} Runtime estimate $\mathcal{T}_{\text{estimate}}$, defined as the product of the peak height and width of $\left| \rmd \tau /\rmd s\right|$, as a function of problem size for $d=2,3$ and $4$. In two dimensions where the slope is $\alpha\approx 1$, no speedup is observed. In $d=3$ and $4$, the slopes are smaller than the classical one, indicating a quantum speedup. While in four dimensions the speedup is full, i.e., quadratic, with a slope of $\alpha\approx 1/2$, in three dimensions it is $\alpha \approx 2/3$.}
   \label{fig:lae}
\end{figure}

To estimate the runtime $\mathcal{T}$ of the adiabatic process involving a hardcore boson hopping on an optical lattice, we numerically calculate the product of the peak height $\mathcal{H}=\max_s \left[V_{01}(s)/g^2(s)\right]$ and peak width \hbox{$\mathcal{W}=s_+-s_-$}, where $s_\pm$ are the points at which the peak is halved\footnote{In the usual adiabatic Grover Hamiltonian where the gap is given by $g(s)=\sqrt{1-4(1-1/N)s(1-s)}$ and the matrix element is $V_{01}(s)=\sqrt{N-1}/[N g(s)]$~\cite{hen:14b}, one can readily calculate the peak height $\mathcal{H}$ and width $\mathcal{W}$ analytically to obtain $\mathcal{T}_{\text{estimate}}=\mathcal{H} \times \mathcal{W}=\sqrt{(4^{1/3}-1)N}$ revealing the correct $\sqrt{N}$ scaling.}.
In fig.~\ref{fig:lae} we show the scaling of $\mathcal{T}_{\text{estimate}}=\mathcal{H} \times \mathcal{W}$ with problem size $N=L^d$ for dimensions $d=2,3$ and $4$, as calculated via exact-numerical diagonalization. 
As the scaling analysis shows, in two dimensions, the obtained slope is $\alpha \approx 1$, similar to the classical slope, indicating no quantum speedup. In (the unphysical case of) four dimensions, the slope is precisely half that of the classical one, indicating a quadratic speedup, consistent with the bounds derived in the previous section. In three dimensions, we find that the slope is $\alpha = 2/3$, denoting a (sub-quadratic) quantum speedup with a runtime that scales as $N^{2/3}$. 

\begin{figure}[htp] 
   \centering
   \includegraphics[width=0.9\columnwidth]{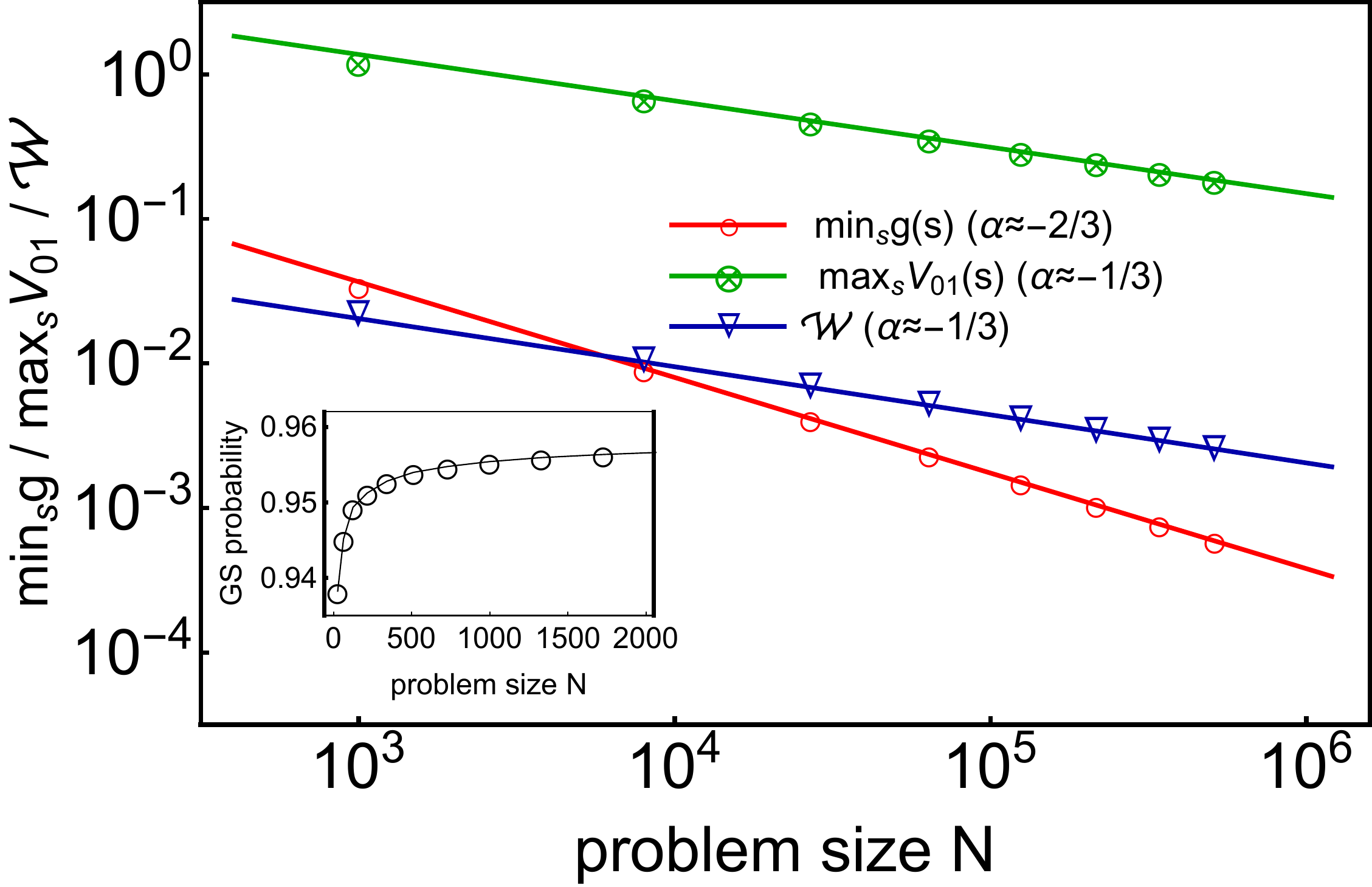} 
   \caption{{\bf Numerical scaling of the minimum gap, the maximum matrix element and width of peak in three dimensions.} While the minimum gap ($\circ$) scales as $N^{-2/3}$, the maximum matrix element ($\otimes$) and the width of the region within which the gap is small ($\nabla$) both scale as $N^{-1/3}$. {\bf Inset:} Ground state probability at the end of a constant-rate adiabatic evolution in three dimensions for runtimes that scale linearly with problem size $N$.}
   \label{fig:d3}
\end{figure}

The quantum speedup observed in the three dimensional case can be attributed to the combined scaling of three factors: the minimum gap, the maximal matrix element and the width of the region over which the two are dominant. The scaling of the three with problem size is  shown in fig.~\ref{fig:d3}. The minimum gap $\min_s g(s)$ scales as $N^{-2/3}$ and over a region of width $\mathcal{W} \sim N^{-1/3}$. The combination of these two factors alone, namely, $\mathcal{W}/\min_s g^2(s)$ which is usually sufficient to determine the scaling of the total runtime for adiabatic processes, yields in this case a combined scaling of $O(N)$. A constant rate adiabatic evolution on the other hand which scales as \hbox{$\max_s V_{01}(s)/\min_s g^2(s)$} as per eq.~(\ref{eq:T}) amounts to an $O(N)$ scaling as well. The ground state probability for a constant-rate adiabatic process with runtimes that scale with problem size $N$ [consistently with the adiabatic condition, eq.~(\ref{eq:T})] is given in the inset of fig.~\ref{fig:d3}. It is the combination of all three factors via which a speedup of $O(N^{2/3})$ is achieved. 

\begin{figure}[htp]
\includegraphics[width=0.45\textwidth]{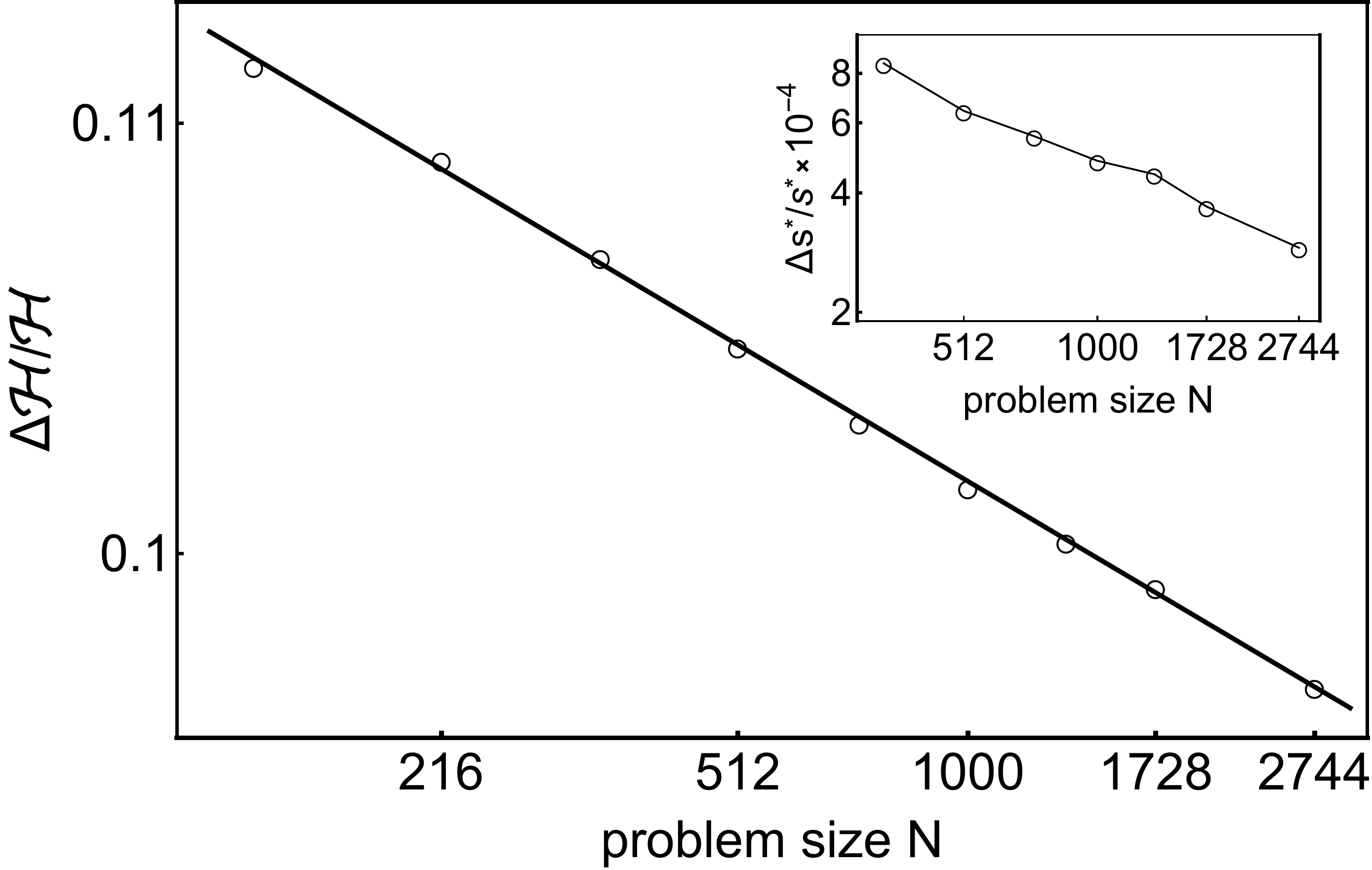}
\caption{{\bf Relative spread of peak height (inset: peak location) of \hbox{$|\rmd \tau/\rmd s|=V_{01}(s)/g^2(s)$} in three dimensions due to open boundary conditions.} The relative spread of both height and location decay with increasing problem size, indicating the diminishing effects of open boundaries on the shape of \hbox{$|\rmd \tau/\rmd s|$.}}
\label{fig:boundary}
\end{figure}

Since three dimensional optical lattices can not in reality support periodic boundary conditions, it is instructive to also analyze the effects of open boundary conditions on the performance of the algorithm. In the analysis presented above, the periodic boundary conditions provided the system with the translational symmetry that allowed for the calculation of the gap and matrix element of the Hamiltonian with the help of which quantum speedup was demonstrated. Open boundaries break the translational symmetry, and the location of the marked site $i_*$ affects accordingly the spectrum of the system Hamiltonian depending on its distance from the boundary. To quantify the effects open boundaries, we measure the changes in \hbox{$|\rmd \tau/\rmd s|=V_{01}(s)/g^2(s)$} for the various choices of marked site $i_*$. We illustrate the diminishing effects of the open boundaries in fig.~\ref{fig:boundary} in which the relative spread of the height and location of the peak of \hbox{$|\rmd \tau/\rmd s|$} is shown as a function of increasing problem size. As one might expect, the effects of the open boundaries become less and less discernible with increasing system size.

\section{Summary and conclusions} We have presented a setup in which a single quantum particle placed in an optical lattice can be used to carry out a computation. Specifically, we have demonstrated that a hardcore boson can adiabatically search through a lattice faster than a classical particle in three dimensions and above. 

That a quantum speedup can already be observed in three dimensions, suggests that the mechanism proposed above is not unrealistic. 
Thanks to recent developments in the field of ultra-cold gases which have matured to a stage where superfluid to Mott-insulator transitions are physically realizable~\cite{jaksch,bloch,Spielman,Spielman2,greiner}, experiments such as the one described above can in principle already be carried out. In practice, the realizability of the above setup depends on several factors. Aside from the open boundary conditions discussed above, foremost is the high level of control over the various parameters of the model, namely the strength of the hopping term, the onsite repulsion and the site-dependent chemical potential, required to carry out the experiment. While the sensitivity of the performance of the algorithm proposed above to errors and imperfections in the setup has not been analyzed here, it is plausible to assume that a certain threshold of errors could still be tolerated to achieve an observable speedup. 

The mere possibility that quantum adiabatic speedups may actually be observed and verified in a lab using currently available resources is of both theoretical and experimental significance. Novel setups such as the one introduced here may pave the way to new avenues of practical quantum computation that have not been considered so far, where quantum particles transitioning from delocalized to localized states are used as resources for speedier calculations of appropriately thought out computational tasks. 

It is interesting to note that a three-dimensional quantum random walker whose dynamics is governed by the same Hamiltonian does not provide a similar speedup~\cite{spatialSearchChilds,wong}. Another interesting question that arises concerns other possible realistic setups in which quantum speedups can be achieved in two-dimensional geometries which are easier to set up experimentally. 
It would be of interest to find additional examples, as well as practical applications, where quantum annealing on optical lattices yields experimentally achievable quantum speedups using present-day technologies. 

\section*{Acknowledgements} 
We thank Tameem Albash, Mohammad Amin, Daniel Lidar, Eleanor Rieffel and Marcos Rigol for useful comments and discussions.

\bibliography{refs}

\end{document}